\documentclass[useAMS,usenatbib,usegraphicx]{mn2e}
\usepackage{epsfig}
\usepackage{amsmath}         %need for line break in equ 3
\usepackage{rotating}           % for sideways tables/figures
\usepackage{color}                % for colour font

\usepackage{times}

\def\kms{km ${\rm s}^{-1}$}

\def\ch2{$\chi^2$}
\def\dg{$^{\circ}$}
\def\Mo{M$_\odot$}
\def\kms {\hbox{${\rm km\ s}^{-1}$}}

\def\scm  {$\hbox{{\rm cm}}^{-2}$}    %cm-2
\def \HI {H{\sc \,i}}
\def \WpHz {W Hz$^{-1}$}

\def\lapp{\ifmmode\stackrel{<}{_{\sim}}\else$\stackrel{<}{_{\sim}}$\fi}
\def\gapp{\ifmmode\stackrel{>}{_{\sim}}\else$\stackrel{>}{_{\sim}}$\fi}

\usepackage{xspace}

\title[Localised \HI\ 21-cm absorption at $z=0.24$]{Localised \HI\ 21-cm absorption towards a double-lobed {\boldmath $z=0.24$} radio galaxy}
\author[S. J. Curran et al.]{S. J. Curran$^{1}$\thanks{E-mail:
sjc@phys.unsw.edu.au}, M. T. Whiting$^{2}$,  J. K. Webb$^{1}$ and  R. Athreya$^{3}$ \\
$^{1}$School of Physics, University of New South Wales, Sydney NSW 2052, Australia\\
$^{2}$CSIRO Australia Telescope National Facility, PO Box 76, Epping NSW 1710, Australia\\
$^{3}$Indian Institute of Science Education and Research, 900, NCL Innovation Park, Dr Homi Bhabha Road
Pune, Maharashtra 411008, India}

\begin{document}

\date{Accepted ---. Received ---; in original form ---}

\pagerange{\pageref{firstpage}--\pageref{lastpage}} \pubyear{2011}

\maketitle

\label{firstpage}

\begin{abstract}
  We present the results of a mini-survey for associated \HI\ 21-cm absorption at $z\leq0.42$ with the Giant Metrewave
  Radio Telescope.  Our targets are radio galaxies, selected on the basis that the $\lambda\approx1216$ \AA\
  luminosities are below $L_{\rm UV}\sim10^{23}$ \WpHz, above which there has never been a detection of 21-cm
  absorption. Of the three sources for which we obtained good data, two are unclassified active galactic nuclei (AGN)
  and one is type-2. Being a non-detection, the type-2 object is consistent with our previous result that 21-cm absorption in radio sources is not
  dictated by unified schemes of AGN.  In the case of the detection, the absorption only occurs towards one of the two
  resolved radio lobes in PKS 1649--062.  If the absorption is due to another intervening galaxy, or cool \HI\ gas in
  the intergalactic medium, covering only the south-west lobe, then, being at the same redshift, this is likely to be
  gravitationally bound to the optical object identified as PKS 1649--062.  If the absorption is due to an inclined disk
  centred between the lobes, intervening the SW lobe while being located behind the NE lobe, by assuming that it
  covers the emission peak at $\approx150$ kpc from the nucleus, we estimate a dynamical mass of
  $\approx3\times10^{12}$~\Mo\ for the disk.

\end{abstract}


\section{Introduction}
\label{intro}
Redshifted \HI\ 21-cm absorption can provide an excellent probe of the contents and nature of the
early Universe, through surveys which are not subject to the same flux and magnitude limitations
suffered by optical studies. In particular, observation of the epoch of re-ionisation  (e.g. \citealt{cgfo04}),
measurement of the contribution of the neutral gas content to the mass density of the Universe
(see \citealt{cur09a}), as well as measuring any putative variations in the values of the
fundamental constants at large look-back times (\citealt{twm+05,tmw+06}).

% it was 41 intervening - see cwm+10 
% cw10 (z > 0.1) gives 32 with another 2 new from cwm+10, so 41 + 32 + 2 = 75
Such absorption is, however, currently rare, with only 75 \HI\ 21-cm absorption systems at
$z\geq0.1$ known -- 41 of which occur in galaxies intervening the sight-lines to more distant
quasars, with the remainder arising in the hosts of radio galaxies and quasars (summarised in
\citealt{cur09a,cw10}, respectively).\footnote{Adding a further two associated absorbers found by
  \citet{cwm+10}.} 
%%%%%%%%%%%%%%%%%%%%%%%%%%%%%%%%%%%%%%%%%%%%%%%%%%%%%%%%%%%%%%%%%%%%%%%%%%%%%%%%%%%%%%%%
% papers/47
 % 22 all-det-klm+09-actual.dat  + 1 of sgp+10
% 24 all-non-klm+09-actual.dat
%%%%%%%%%%%%%%%%%%%%%%%%%%%%%%%%%%%%%%%%%%%%%%%%%%%%%%%%%%%%%%%%%%%%%%%%%%%%%%%%%%%%%%%%
While the 50\% detection rate in intervening absorbers can be attributed to flux coverage
effects, introduced by the geometry of a flat expanding Universe \citep{cw06,ctd+09}, the detection rate in these
latter ``associated'' systems is currently attributed to unified schemes of active galactic nuclei
(AGN), where only type-2 objects present a dense column of absorbing gas along our sight-line. We
\citep{cww+08} have however recently found that the ultra-violet luminosity of the AGN is the key in
determining whether 21-cm absorption is detected, with AGN type having little bearing on this,
contrary to the current consensus (\citealt{gs06a} and references therein).  Indeed, UV luminosities
may also be the root cause of other observed properties, seemingly responsible for the 21-cm
detection rate in the hosts of radio galaxies and quasars (\citealt{cw10} and verified by \citealt{gd11}).

We hence suggest that the paucity of redshifted 21-cm absorption is
due to the optical selection of targets, where at such large distances
only the most UV luminous are known. We have therefore embarked on
several observing campaigns, at various redshifts, of optically faint
sources in order to test this hypothesis and increase the number of
associated 21-cm absorbers known at $z\gapp0.1$.  In this letter we
present the first of these, a mini-survey using the 1000--1450 MHz band
of the Giant Metrewave Radio Telescope (GMRT), in which we find 21-cm
absorption towards one of the two lobes resolved in PKS 1649--062, a $z=0.24$ radio
galaxy \citep{brl99}.

\section{The 1420 MHz band GMRT mini-survey} %observations

\subsection{Source selection}

As per our previous surveys \citep{cwm+06,cww+08,cwm+10}, we selected targets from the Parkes
Half-Jansky Flat-spectrum Sample (PHFS, \citealt{dwf+97}), a source of bright and generally compact
radio objects for which there exists comprehensive optical photometry \citep{fww00}. In order to
ensure that the $\lambda\approx1216$ \AA\ luminosities of our targets are below the critical $L_{\rm
  UV}\sim10^{23}$ \WpHz, above which 21-cm has never been detected \citep{cww08}, we target those
which are sufficiently faint at a given redshift. However, applying the $B-z$ curve (figure 5 of
\citealt{cww09}) as an initial diagnostic, yields only two PHFS sources with $B\gapp22$ in the
307--347 MHz band ($z=3.09 - 3.63$), both of which have been previously searched for 21-cm
\citep{cww+08}. Of the other available bands, there are three with $B\gapp20$ in the 580--640 MHz
band ($z=1.22 - 1.45$), whereas in the 1000--1450 MHz band ($z\leq0.42$) there are 15 sources with
measured redshifts and $B\gapp19$.\footnote{Although $B\gapp17$ should be close to ensuring $L_{\rm
    UV}\lapp10^{23}$ \WpHz\ at these redshifts
  \citep{cww09}.} % ./21cm-bands IN proposals/2009/dim_quasars (not C/astro) ON PHFS-B-z_final.dat

We therefore selected the ten faintest in this band, for which we also requested time to search for OH 18-cm absorption, on the basis of their
large blue--near-infrared colours (all ten have $B-K\geq5.33$, cf. $\gapp6$ for the five known redshifted OH absorbers, see \citealt{cwm+10}).
However, time was only awarded to search for 21-cm in four objects and so we selected the two reddest which had not been
previously searched, PKS 0036--216 ($B-K=6.22$) and 1128--047 ($B-K=7.55$), as well as the two with the largest flux densities (estimated to be $\sim2$ Jy),
0153--410 and 1649--062.

\subsection{Observations and data reduction}

\begin{figure*}
%\centering \includegraphics[angle=0,scale=0.35]{map+spectra.ps} 
\centering \includegraphics[angle=0,scale=0.35]{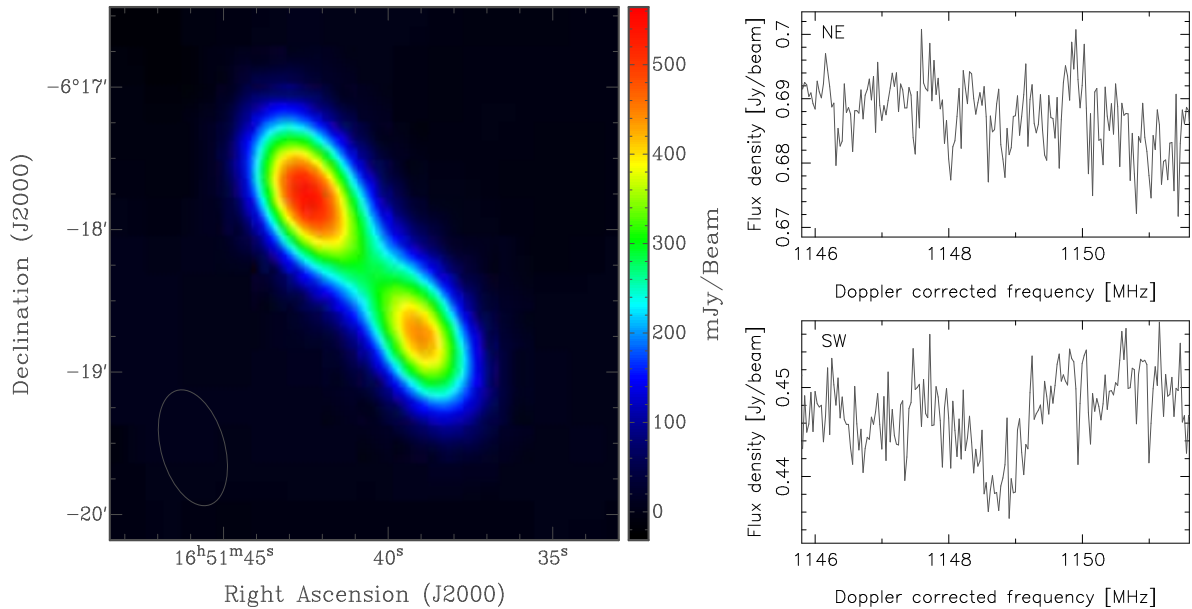} %astro-ph not accepting ps files !!!
% ALOS HAVE map+spectra_mJy+Jy.ps IN WHICH SPECTRA ARE LABELLED WITH Jy AND map+spectra.ps WITH INTEGRATED FLUX FOR IMAGE
\caption{Left: The GMRT 1149 MHz continuum image of PKS 1649--062 (cf. the 408 MHz image of \citealt{brl99},
which also resolves the two features). The
  beam is shown in the bottom left corner. Right: The spectra extracted from the north-east and
  south-west lobes of the image, shown at the observed channel width of 8.2 \kms. } 
\label{spectra+map}
\end{figure*}

The four shortlisted targets (listed in Table \ref{sum}) were observed with the 1000--1450 MHz receiver
backed with the FX correlator over a bandwidth of 8 MHz split into 256 channels. This setup was chosen in order to give
a channel spacing of $\approx8$ \kms,
%(cf. a FWHM of 7 to 370 \kms\ for the known associated 21-cm absorbers, see \citealt{cw10}), 
while maintaining a coverage of $\Delta z \approx
\pm0.004$, thus covering any
uncertainties in the redshifts of the targets. Regarding the observations of these:\\
{\bf PKS 0036--216} was observed at a central frequency of 1061 MHz for a total of 0.90 hours on 26
November 2009. The delays were self calibrated, with PKS 0023--26 being used for the bandpass
calibration. Minimal radio frequency interference (RFI) meant that only non-functioning antennas had
to be flagged, leaving 365 good baseline pairs. For this, and the other targets, the data were
reduced using the {\sc miriad} interferometry reduction package, from which a spectrum was extracted
from the cube.
The source was unresolved by the $4.8"\times3.6"$ synthesised beam.\\
{\bf PKS 0153--410} was observed at a central frequency of 1159 MHz for a total 1.82 hours on 26 November
2009. Again, the delays were self calibrated, with 3C\,48 being used for bandpass
calibration. Severe RFI over the length of the observation meant that no image could be
produced.\\ %, nor even an average of the baseline pairs.\\
{\bf PKS 1128--047} was observed at a central frequency of 1122 MHz for a total of 0.93 hours on 25 July
2010.  The delays were calibrated using LBQS 1148--0007 and the bandpass using 3C\,286. RFI was
minimal, leaving 365 good baseline pairs. The source was unresolved by the $5.6"\times3.2"$ synthesised beam.\\
{\bf PKS 1649--062} was observed at a central frequency of 1149 MHz for a total of 2.29 hours on 27
November 2009. The delays were calibrated using PKS 1621--11, with bandpass calibration by 3C\,273
and 3C\,468. Although there was little sign of RFI, after calibration, using either bandpass
calibrator, some of the antenna pairs (involving mostly antenna number 15 and above)
exhibited flux densities which were so low as to be comparable with the r.m.s. noise and the inclusion of these 
prevented the production of any reasonable image. We therefore used a script to automatically flag all of those with
a mean flux-to-noise ratio below some specified value. Only once this value reached 10 was a
reasonable image (that shown in Fig. \ref{spectra+map}, left) obtained and since this left us with
only 60 baseline pairs, further flagging would have been detrimental to this. 
% b1649-062.1149.cal-NE.dat.ps & b1649-062.1149.cal-SW_2.dat.ps PRODUCED FROM spectrum.c 
% Manual axis values in GMRT_2010.log   i.e.               ./spectrum < SW.in      ./spectrum < NE.in
%THEN INKSCAPE into map+spectra.svg
The flagging of the 
longer baselines meant that the synthesised beam was much coarser than
for the other observations ($50"\times29"$). Nevertheless, this was sufficient to resolve a double lobed radio
source, in which 21-cm absorption was detected towards
one of the lobes. % (Fig .\ref{spectra+map}).

\section{Results and discussion}

\subsection{Survey results}
\label{sr}

In Table \ref{sum} we summarise the results of the survey, where we detected 21-cm in one of the three sources
for which there are useful data (see Fig. \ref{spectra+map}, right).
\begin{table*} 
\centering
%\begin{minipage}{177mm}
\begin{minipage}{170mm}
\caption{Summary of the search for 21-cm absorption in the hosts of optically faint PHFS sources which fall into the
1000--1450 MHz band. $\nu_{\rm obs}$ is the observed central frequency [MHz], $B$ is the blue magnitude used
in the selection of the sources, $L_{\rm UV}$ is the $\lambda=1216$ \AA\ luminosity of the target [\WpHz]  
(calculated using the method prescribed in \citealt{cww+08}), $\sigma_{{\rm rms}}$ is the r.m.s. noise [mJy/beam] reached
per $\Delta v$ channel [\kms], $S_{\rm cont}$ is the continuum flux [Jy], $\tau$ is the peak optical depth of the line, where
$\tau=-\ln (1-3\sigma_{{\rm rms}}/S_{\rm cont})$ is quoted for the
non-detections, $N_{\rm HI}$ is the resulting column density [\scm], where $T_{\rm s}/f$
is the spin temperature/covering factor degeneracy of the of 21-cm absorbing gas, followed
by the redshift range over which this applies. Finally, we give the AGN type.}% derived from the corresponding reference.}
\begin{tabular}{@{}r l  c  c  c  c  c  c   r r  c  c  c @{}} 
\hline
PKS &  $z_{\rm em}$ & $\nu_{\rm obs}$  & $B$  & $\log_{10}L_{\rm UV}$ &   $\sigma_{{\rm rms}}$ & $\Delta v$ &$S_{\rm cont}$ & $\tau$ & $N_{\rm HI}$  &  $z$-range & Type \\ %& Ref \\
\hline
0036--216  & 0.338  &   1061 & 21.20 & 19.24  & 2.3 & 8.8 & 0.67 & $<0.010$ & $<1.7\times10^{17}.\,(T_{\rm s}/f)$  & 0.3325--0.3425&   --\\ %& \\  %galaxy 
0153--410  &  0.226 &  1159  & 19.81 & 19.67  &  --- &  8.1 & --- &\multicolumn{2}{c}{\sc  RFI dominated }& --- & 2  \\ %& \\  %Sy2 - NED
1128--047  &  0.266 &  1122  & 21.41 & 19.86  &  4.1 &  8.4 & 0.31 & $<0.040$ &  $<6.1\times10^{17}.\,(T_{\rm s}/f)$  & 0.2615--0.2705& 2  \\ %& \\%Sy? - NED
 % \multicolumn{6}{c}{\line} & 1649--062 &   multicolumn{6}{c}{\line}\\
    1649--062 & \multicolumn{12}{c}{} \\
 NE&     0.236  &   1149  & 20.44 & 18.91 &  5.1  & 8.2 &  0.69 &  $<0.022$ &  $<3.3\times10^{17}.\,(T_{\rm s}/f)$  &   0.2319--0.2395 &  -- &\\ % B99\\ %radio galaxy - NED
    SW                & ...          &...          &...          &...          & 4.1  & ...  &  0.45   & 0.029 &  $8\pm2\times10^{18}.\,(T_{\rm s}/f)$  &  0.23647 & ...& \\ % ...  \\
\hline       
% B=20.44 is from the USNO catalogue 
\end{tabular}
%{\flushleft References: B99 -- \citet{brl99} {\color{red}{\bf OTHER REFERENCES?}}}
\label{sum}  
\end{minipage}
\end{table*} 
 \begin{figure}
%\centering \includegraphics[angle=270,scale=0.70]{2-N-L.eps}  %UPDATE WITH NEW RESULTS
\centering \includegraphics[angle=270,scale=0.72]{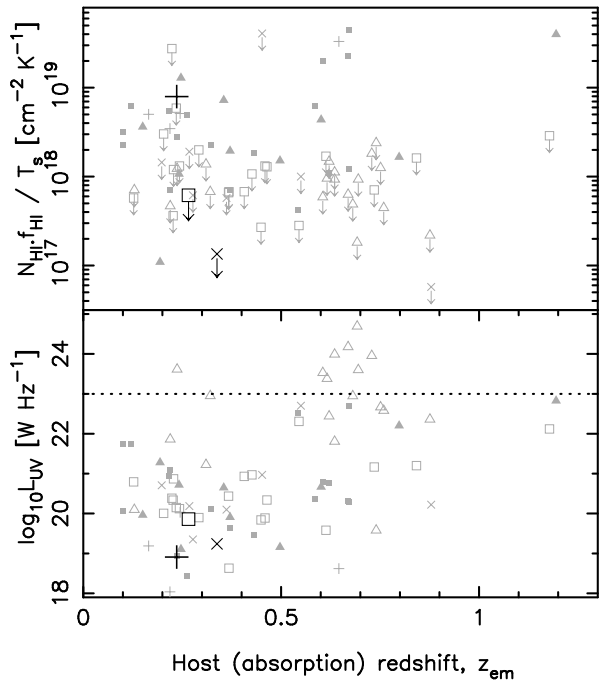}
\caption{The scaled velocity integrated optical depth of the \HI\ line
  ($1.823\times10^{18}.\int \tau dv$) [top] and the ultra-violet
($\lambda\approx1216$ \AA) luminosity [bottom] versus the host redshift for the
$z\geq0.1$ radio galaxies and quasars searched in associated 21-cm
absorption.  The filled symbols represent the 21-cm
detections and the unfilled symbols the
non-detections, with the large black symbols designating the new
results presented here.
The shapes represent the AGN classifications, with
triangles representing type-1 objects and squares type-2s ({\bf +} and
{\sf x} designate an undetermined AGN type for a detection and
non-detection, respectively). Limited to $z\lapp1.2$ to show the new results. See \citet{cwm+10}
for the full redshift range. } 
\label{2-N-L}
\end{figure}
%   sort -d -k 8 low-lum-det.txt > temp           5 unclass, 11 type-1, 18 type-2  WILL NOW BECOME 6 unclass, 11 type-1, 17 type-2 
%sort -d -k 8 low-lum-non.txt > temp             9 unclass, 12 type-1,  21 type-2       AND THIS WILL BE 20 type-2 so now 17 out of 38
Showing the limits obtained in the top panel of Fig.~\ref{2-N-L}, we see that the column density of the detection is in the
nominal range of those previously searched and that the two non-detections are at comparatively deep limits. In the bottom panel we show how the $\lambda=1216$~\AA\
luminosities
of our targets compare to those previously searched for 21-cm absorption, where, as expected from our source
selection, these are well below $L_{\rm UV}\sim10^{23}$ \WpHz, at the lower end of the searched UV range.

As discussed in Sect. \ref{intro}, the detection of 21-cm absorption in radio galaxies and quasars is usually
attributed to unified schemes of AGN, where type-2 objects generally exhibit absorption, whereas type-1
objects do not. However, once the effect of the UV luminosity is removed, \citet{cww+08} found that there is a near
50\% detection rate for {\em both} type-1 and type-2 objects. Unfortunately, two of the targets cannot be classified;
1649--062 does not have an AGN classification as its optical spectrum is purely that of an elliptical galaxy \citep{brl99},
although it does have a \citet{fr74} type-{\sc ii} radio morphology,
while 0036--216 does not have any published spectrum suitable for discerning its AGN type.
The one non-detection for which there exists a classification (1128--0047, Table~\ref{sum}) is a type-2 object,
which would be expected to exhibit absorption, where this determined by the unified schemes. Adding this
to the statistics of the  $z\geq0.1$ searches, takes the detection rates to 11 out
of 23 (48\%) and 
17 out of 38 (45\%) for the type-1 and type-2 AGN, respectively.

\subsection{The absorption in PKS 1649--062}
\label{orig}

\begin{figure}
\centering \includegraphics[angle=270,scale=0.32]{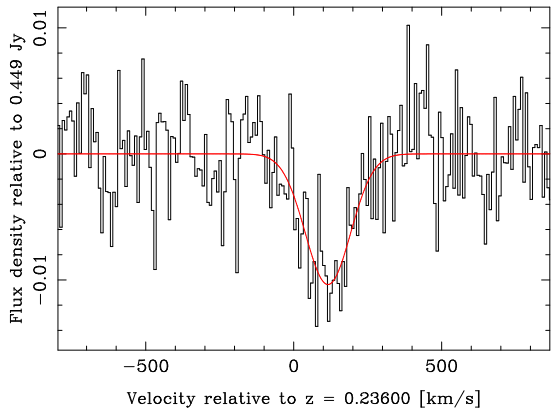}
\caption{Gaussian fit to the absorption towards the SW lobe. } 
\label{gauss}
\end{figure}

We believe that this is the highest redshift detection of \HI\ 21-cm absorption against an extended  radio lobe to date,
the first cases being 3C\,234 ($z= 0.185$, \citealt{pih01})
and Coma A ($z=0.086$, \citealt{mot+02}).
%Regarding the absorption detected here, 
In Fig. \ref{gauss} we show the Gaussian fit to the absorption profile of 1649--062. 
The full-width half maximum (FWHM) of 181 \kms\  gives a velocity integrated line
strength of 1.91 Jy \kms, which agrees well with the $2.0\pm0.5$ Jy \kms\ obtained
from summing the channels.
The peak absorption occurs at a velocity offset of 116 \kms\ from the reference $z=0.236$ (with a depth of
  10 mJy), which gives a redshift of $z=0.23648$. This is consistent, within uncertainties, with the $z = 0.236\pm0.002$ obtained 
from the optical spectrum \citep{brl99}.
 From the image, the centre of the NE and SW emission regions are separated by $\approx80"$ (both $\approx40"$ from the
point of minimum flux), 
which gives a projected extent of $\approx300$ kpc spanning between both emission peaks.\footnote{For $H_{0}=71$~km~s$^{-1}$~Mpc$^{-1}$, $\Omega_{\rm matter}=0.27$ and
$\Omega_{\Lambda}=0.73$.}

\subsubsection{Associated absorption by a large galactic disk}
\label{lgd}

If the 21-cm absorption in PKS 1649--062 arises in a disk centred between the two radio lobes,  it must have a non-negligible inclination along our sight-line in order
to absorb in the SW, where it would occult the lobe, while being behind the NE lobe. For the disk to cover
the peak emission at the SW lobe, its radius, $r_{\rm disk}$,  rotational velocity, $v_{\rm disk}$, and
inclination, $i$, would be related to the observed properties via 
\begin{equation}
 %v_{\rm disk}\geq \frac{r_{\rm disk}}{r_{\rm lobe}} \,v_{\rm obs}\,\tan\,i,
r_{\rm disk}\geq\left(\frac{v_{\rm disk}}{v_{\rm obs}}\right)\,\left(\frac{r_{\rm lobe}}{\tan\,i}\right),
\label{equ1}
\end{equation}
where $r_{\rm lobe}$ is the projected distance of the emission peak from the nucleus and $v_{\rm obs}$ is the
observed rotational velocity. Although we do not have the information to break
the degeneracy, combining equation~\ref{equ1} with Newton's second law gives for the dynamical mass, in solar masses, 
\begin{equation}
M_{\rm dyn} \geq 231\,\left(\frac{v_{\rm obs}}{r_{\rm lobe}}\right)^2 \,r_{\rm disk}^{~~~~~3}\,\tan^2i,
\label{equ2}
\end{equation}
where $r_{\rm disk}$ [pc] is the disk radius rotating at $v_{\rm disk}$ [\kms]. 

\begin{figure}
\centering \includegraphics[angle=270,scale=0.72]{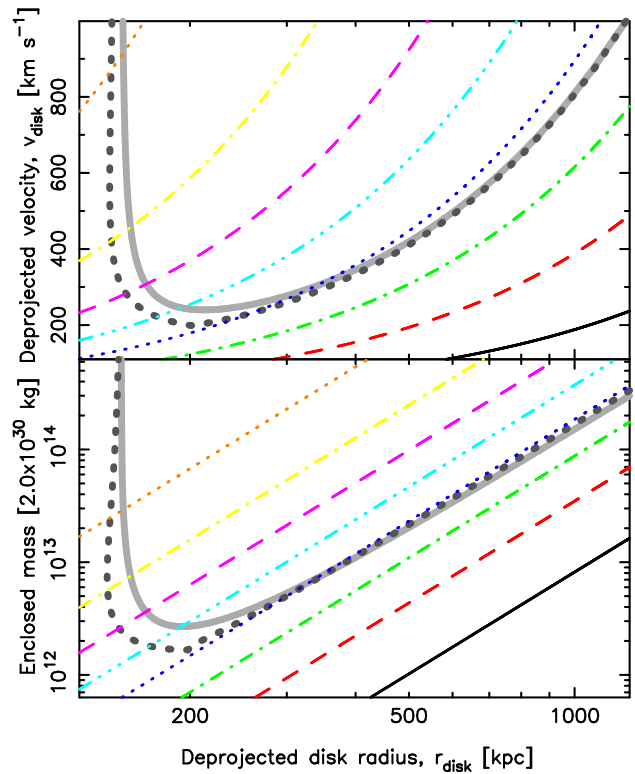} 
\caption{The deprojected rotation velocity (top) and dynamical mass (bottom) versus radius for a disk with
an observed maximum rotational velocity of 160 \kms\ and a projected radius of 150 kpc for various inclinations.
The inclination angles decrease from left to right in steps of 10 degrees, starting at $i=80$\dg\ (dotted curve) and
ending on  $i=10$\dg\ (full curve). The thick contours show the permitted values for the putative disk
at various inclinations (full for a thin disk and broken for a thick disk, see main text).}
% The hatched areas show the permitted regions for $r_{\rm disk}\geq r_{\rm lobe} = 150$~kpc
% and $v_{\rm disk}\geq v_{\rm obs} = 160$~\kms. {\bf HAVE STOPPED THIS AT 300 \kms BUT SHOULD FIND IN LITERATURE
% WHAT MAX ROTATION VELOCITY IS FOR A GALACTIC DISK. ALSO UPPER LIMIT ON RADIUS THEN CAN PUT LIMIT
% ON RIGHT HAND SIDE OF HATCHING}} 
\label{2-inc}
\end{figure}

Applying $r_{\rm lobe}\approx150$ kpc
and the full observed line width of $v_{\rm obs}\approx\pm160$ \kms, in Fig.~\ref{2-inc} we show the results of 
equations \ref{equ1} and \ref{equ2} for various disk inclinations,
which are overlaid with the permitted velocity, radius and
subsequent dynamical mass at various inclinations. These are constrained from
$r_{\rm disk} = r_{\rm lobe}/\sin\,i$ and $v_{\rm disk} = (r_{\rm disk}/r_{\rm em})\,(v_{\rm obs}/\cos\,i)$,
where $r_{\rm em}$ is the extent of the background emission behind the disk. We find this to 
be $\approx200$ kpc (beyond which the summed continuum and absorption profile disappear). The additional factor of $r_{\rm disk}/r_{\rm em}$ is 
used to account for the 
possibility that the observed velocity, in addition to being projected in inclination, may be a projection
due to the background emission not reaching as far as the disk edge, thus not allowing this to absorb. 
That is, this correction accounts for the possibility that the 21-cm absorption may not extend as far as the
21-cm emission from the disk, could this be detected.
Note that the correction is responsible for the climb in velocity above $r_{\rm disk} \approx 200$ kpc 
in Fig.~\ref{2-inc}, where otherwise this would decline to close to the observed velocity by a radius of $\gapp1$ Mpc (where
$v_{\rm disk} \lapp162$ \kms\ and $i\lapp9$\dg).

From this model, we find a minimum mass
of $2.7\times10^{12}$~\Mo\ (for $i\approx50$\dg, where  $v_{\rm
  disk}\approx 240$ \kms\ and $r_{\rm
  disk}\approx 200$ kpc). Without a maximum deprojected velocity we cannot estimate
the maximum possible dynamical mass, but adopting the $\pm290$ \kms,
which is the largest profile
width found in the low redshift ($z\lapp0.04$) \HI\ Parkes All-Sky Survey \citep{ksk+04}, gives a dynamical mass of
$3.3\times10^{12}$~\Mo\ within a 170 kpc radius ($i\approx60$\dg, from the $r_{\rm disk}\leq r_{\rm em}$ part of the curve).

Large (100 kpc-scale) disks have been observed in near-by radio galaxies (e.g. \citealt{mot+02,emo+08}) and,
in order to account for the asymmetric properties observed in some
radio galaxies, ``superdisks'' of gas and dust around the elliptical hosts of powerful radio galaxies,  have been proposed.
These are oriented roughly orthogonal to the radio jets, where one lobe is obscured by the disk \citep{akmv98,gw00}.
Such disks have diameters of at least 75
kpc, although the maximum in the sample of \citet{gw00} is 137 kpc, which is considerably lower than the minimum
estimated diameter of $\approx340$ kpc for the putative disk in PKS 1649--062.  These superdisks can be one third as
thick as they are long and broadening the disk to this ratio with the addition of a $(t_{\rm disk}/2)\,\cos\,i$ term,
where $t_{\rm disk}$ is the disk thickness, and assuming no shear across the edge of the disk (Fig. \ref{2-inc}, broken
contours), still gives a diameter of $\gapp300$~kpc ($i\lapp56$\dg\ and $M_{\rm dyn}\approx 2.7\times10^{12}$~\Mo, for
$r_{\rm disk}\leq r_{\rm em}$) for $v_{\rm disk}\lapp290$~\kms.

\subsubsection{Other possible sources of absorption}

Such a large implied size for a single absorbing disk can be avoided by having the absorbing gas:
%\begin{list}
\begin{enumerate}
  \item {\em Within or close to a neighbouring galaxy}. Referring to an optical image of the field
(Fig. \ref{overlay}), there are at least three candidates for the location of the absorbing gas  lying within the SW radio lobe ({\sf A}, {\sf B} and {\sf C}),
\begin{figure}
\centering \includegraphics[angle=0,scale=0.38]{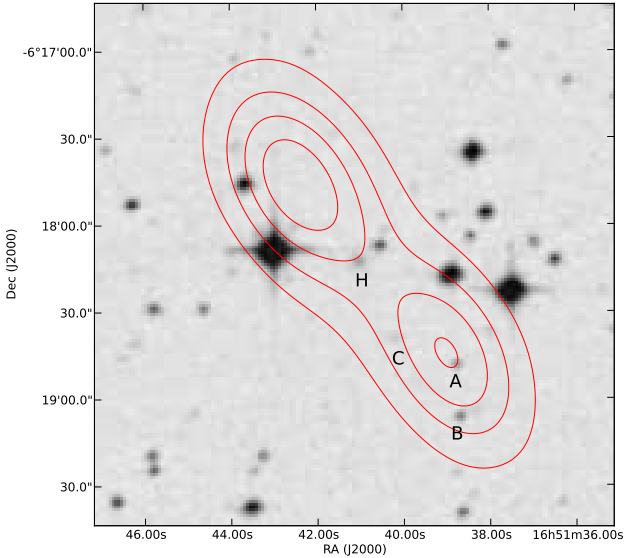}
\caption{The 1149MHz continuum
emission overlayed on a red POSS2 optical image. The host galaxy, as determined by \citet{brl99}, is labelled {\sf  'H'}, with
three potential hosts for the absorbing gas labelled {\sf  'A'} , {\sf  'B'} and {\sf  'C'} (see main text for description).}
\label{overlay}
\end{figure}
although the absorption redshift would require the gas to be located in the same cluster as the host, {\sf 'H'}, which is the only
galaxy in this field with a known redshift \citep{brl99}.
  We list the photometry of the host and the three candidate
objects in Table~\ref{poss2} using the USNO-B survey \citep{mlc+03}, from which we see that the colours ($B-R$) of {\sf
  'A'} and {\sf 'B'} are not as red as the host galaxy. Without spectroscopy, however, it is unclear whether this
difference is due to a difference in redshift, or whether they lack the $4000$\AA\ break of the host galaxy \citep{brl99}.
\begin{table} 
\caption{The $B$ and $R$ magnitudes of the possible absorption features.}
% \begin{tabular}{@{}l  c  c  c @{}} 
% \hline
% Feature & $B$ & $R$ & $B-R$\\
% \hline
% {\sf  H} & 20.44 & 17.43 & 3.01 \\
% {\sf A} & 20.13 & 18.75 & 1.38 \\
% {\sf B} & 19.76 & 17.95 & 1.81 \\
% {\sf C} & ---   & 19.67 & --- \\
% \hline  
\begin{tabular}{@{}l  c  c  c c | l c c c @{}} 
\hline
  & $B$ & $R$ & $B-R$ & \vline  &  & $B$ & $R$ & $B-R$\\
\hline
{\sf  H} & 20.44 & 17.43 & 3.01 & \vline& {\sf B} & 19.76 & 17.95 & 1.81 \\ %{\sf A} & 20.13 & 18.75 & 1.38 \\
%{\sf B} & 19.76 & 17.95 & 1.81 & \vline &  {\sf C} & ---   & 19.67 & --- \\
{\sf A} & 20.13 & 18.75 & 1.38& \vline &   {\sf C} & ---   & 19.67 & --- \\
\hline  
\end{tabular}
\label{poss2}
\end{table}

\item {\em Arising from tidal debris left over from a major merger event}. The presence of 21-cm absorption as 
far as $\approx150$ kpc from the nucleus may be naturally explained by the cold gas left over from a collision, 
a common mechanism in the formation of powerful radio galaxies \citep{hsb+86,bhv92}. The component of this debris which remains gravitationally bound
may eventually form a large \HI\ disk structure on the time-scale of a Gyr (e.g. \citealt{bar02,mot+02,emt+06}).

\item {\em Cool gas in the intergalactic medium (IGM)}. \citet{pih01} describes a similar situation to that of PKS 1649--062,
where 21-cm absorption is detected primarily towards one of the two lobes in 3C\,234. This is interpreted as cool gas
in the IGM, possibly where the radio lobe interacts with this, and so may also explain the localised absorption we observe.

\end{enumerate}
%\end{list}

% In summary, the disk model relies on several assumptions, stated in Sect. \ref{lgd}, whereas the ``unrelated'' galaxy
% must be part of the same cluster as the optical source identified as PKS 1649--062.\footnote{Somewhat akin to the 21-cm
%   absorption in the complex sight-line towards 3C\,336 \citep{ctm+07}.} Optical spectroscopy of the other three galaxies
% in the SW lobe could give redshifts with which to determine whether any of these could be the source of the absorption
% and future Square Kilometre Array observations of the 21-cm emission and absorption in this field should shed light on
% this intricate absorption complex.

%\newpage
\section*{Acknowledgements}

We would like to thank Anant Tanna for the digitised 21-cm 
spectrum of  NGC\,0134 from \citet{ksk+04}.
This research has made use of the NASA/IPAC Extragalactic Database
(NED) which is operated by the Jet Propulsion Laboratory, California
Institute of Technology, under contract with the National Aeronautics
and Space Administration. This research has also made use of NASA's
Astrophysics Data System Bibliographic Services. 

%\bibliographystyle{apj}
%\bibliographystyle{mn2e}
%\bibliography{aa,ref}

\label{lastpage}

\end{document}